\documentclass{article}
\usepackage{graphicx} 
\usepackage[round]{natbib}
\usepackage{longtable}
\usepackage[utf8]{inputenc}
\usepackage{amsmath}
\usepackage{amsfonts}
\usepackage{graphicx}
\graphicspath{ {./images/} }
\usepackage[dvipsnames]{xcolor}

\providecommand{\keywords}[1]
{
  \small	
  \textbf{\textit{Keywords---}} #1
}
\newcommand{\overbar}[1]{\mkern 1.5mu\overline{\mkern-1.5mu#1\mkern-1.5mu}\mkern 1.5mu}

\title{New accessibility measures based on unconventional big data sources}
\author{Arbia, G., Nardelli, V., Salvini, N. and Valentini, I.}
\date{}

\begin{document}

\maketitle

\begin{abstract}
In health econometric studies we are often interested in quantifying aspects related to the accessibility to medical infrastructures. The increasing availability of data automatically collected through unconventional sources (such as webscraping, crowdsourcing or internet of things) recently opened previously unconceivable opportunities to researchers interested in measuring accessibility and to use it as a tool for real-time monitoring, surveillance and health policies definition. This paper contributes to this strand of literature proposing new accessibility measures that can be continuously feeded by automatic data collection. We present new measures of accessibility and we illustrate their use to study the territorial impact of supply-side shocks of health facilities. We also illustrate the potential of our proposal with a case study based on a huge set of data (related to the Emergency Departments in Milan, Italy) that have been webscraped for the purpose of this paper every 5 minutes since November 2021 to March 2022, amounting to approximately 5 million observations.

\end{abstract}\hspace{10pt}


\section{Introduction}

The study of the spatial accessibility to infrastructures has long been the concern of geographers, urban planners and regional economists among the others (see e. g. \cite{hansen_1959_how}). Generally speaking, spatial accessibility measures originate from a spatial econometric framework based on the interaction of 3 major factors \citep{luo_2003_measures}, namely: (i) supply (the geolocated infrastructure), (ii) demand (referring to the individuals who are likely to use the own infrastructure), and (iii) mobility costs from demand locations to supply locations. Further elements, such as distance decay functions and threshold trip time, are also often introduced to better capture people's willingness to visit infrastructures.

In particular, accessibility analysis has proved precious in the study of health facilities, for instance to identify areas characterized by lower/higher accessibility or to emphasize the role of community facilities (i. e. hospitals or Emergency Department). Various spatial methods have been used to quantify the concept of accessibility including the general class of gravity models \citep{isard1954location, hansen_1959_how, wilson1976statistical} and the two-step floating catchment area method (2SFCA \citep{khan_1992_an, luo_2003_measures}. GIS tools have also been largely employed to monitor health phenomena, to evaluate the socio-economic consequences of policies, to determine optimal location of services and to disentangle the relationships between accessibility disparities and health outcomes \citep{mclafferty_2003_gis, higgs_2004_a, neutens_2015_accessibility}. In this realm of research spatial accessibility measurements are becoming more and more sophisticated due to the wider availability of automatically collected data (such as, e. g., those collected through Internet of Things or electric medical devices) and of more powerful computational tools. 


The present paper moves within this general framework aiming to suggest new accessibility indicators that can be calculated in real-time using webscraped datasets.
The rest of the paper is organized as follows. Section 2 is devoted to a review of the various accessibility measures found in the literature. Section 3 introduces new accessibility indicators based on the availability of real-time streaming of data and illustrates their practical use in order to measure policy impact and to explain dynamics. Section 4 reports the results of an empirical analysis focused on the study of spatial accessibility to Emergency Departments (ED) within the Milanese metropolitan area. Finally Section 5 concludes.

\section{A review of the traditional accessibility indices} \label{review}

Access in the healthcare context could be defined as a patient's ability to obtain the correct treatment, from the right provider in a certain space and time \citep{saurman_2016_improving, joyce_2018_use}.
The idea of "health accessibility" encompasses the challenges that must be addressed in order to receive necessary medical care. These issues concern both the production of health services (e.g. number of hospitals, geographic spread, etc.) as well as the demand for them.
A geographical analysis of accessibility is typically directed to suggest possible explanations to the unequal distribution of health services \citep{weiss_2018_a} and to the identification of areas characterized by only a limited accessibility so as to be able to direct political decisions and resources allocation \citep{Mougeot2018, kang_2020_rapidly, Balia2020} aiming at removing spatial barriers between patients and providers that limit the effects of healthcare and preventive service and are often responsible of poorer health outcomes \citep{neutens_2015_accessibility}.

While in this literature the Gravity Model (GM) paradigm was largely dominant until 20 years ago \citep{stewart1941,isard1954location,hansen_1959_how, wilson1976statistical}, in the early 2000s the 2SFCA  method \citep{khan_1992_an, luo_2003_measures, park_2021_a} and the Rational Agent Access Models \citep{saxon2020, saxon2021} emerged as possible alternatives \citep{luo_2003_measures},
see also \cite{shen_1998_location} and \cite{khan_1992_an}. Accessibility is also studied through spatial interaction models but they are specified to be applied to flows data (e.g. agents, goods between points or regions). It is commonly agreed by practitioners in regional science, planning, demography, and economics that gravity models with a Newtonian designation are still among the most widely used analytical tools to study the interaction between agents observed in space and time.

The gravity model in the accessibility context counts the number of opportunities (i.e., supply facilities) available from a particular position while accounting for spatial penalisation \citep{hansen_1959_how} and assuming a uniform distribution of individuals across spatial units. The model can be formalized as follows:

\begin{equation}\label{shen}
    A_i = \sum_{j} \frac{S_j f(d_{ij}))}{\sum_k D_k f(d_{kj}))}
\end{equation}

with $A_i$ denoting accessibility at location $i$, $S_j$ is the weight of supply facility at location $j$, and $d_{ij}$ denotes the distance penalisation between location $i$ and location $j$.  The function $f(.)$ represents a decay function, generally quadratic.
The number of individuals (i.e., demand) at location $k$ is denoted with the symbol $D_k$.

The 2SFCA technique tries to solve the gravity's model constraints, when each supply facility is assumed to offer service to every demand site in the case of an inadequate distance decay function \citep{luo_2003_measures, wang_2020_from}. Efforts have been dedicated to quantify spatial accessibility through this technique \citep{YANG2021113656, JIA2022115458, MULLACHERY2022115307} A threshold time is used to reflect the customer's maximum travel stretch and to specify the areas accessible within the threshold travel time. We call this a "catchment area". This approach measures spatial accessibility in two steps: through a forward pass (from $A_j$ to $R_j$, as reported  in Equation \ref{first-step}) and through a backward pass (from $R_j$ to $A_j$, as presented in Equation  \ref{second-step}) with $A_j$ represnting accessibility and $R_j$ representing reachability. Formally we define:

\begin{equation}\label{first-step}
    R_j = \frac{S_j}{\sum_{k \in \{ d_{ij} \leq d_0  \}} D_k f(d_{kj}) }
\end{equation}

\begin{equation}\label{second-step}
A_j = \sum_{j \in \{d_{ij} \leq d_0\}} R_j f(d_{ij}) = \sum_{j \in  \{d_ij \leq d_0\}} \frac{S_j f(d_{i,j}))}{\sum_{k \in \{ d_{ij} \leq d_0  \}} D_k f(d_{kj}) }
\end{equation}

Where $R_j$ denotes the supply-to-demand ratio of the supply facility at location $j$. The weight of the supply facility at point $j$ is denoted by $S_j$. $D_k$ represents the demand (e.g., population) at k; $d_{kj}$ or $d_{ij}$ is the cost of traveling from point $k$ (or $i$) to point $j$; $f(.)$ is an penalisation function; $d_0$ is the threshold travel cost; and $A_i$ is the accessibility measure at site $i$. Thus the 2SFCA technique develops in two phases. First of all the supply-to-demand ratio of each supply facility is computed (see Equation \ref{first-step}); the weight of the supply facility is divided by the demand's total, which determines each of which sites fall within the catchment area (i.e., it is accessible within the maximum admitted travel time). Then the supply-to-demand ratio of supply facilities is topped up in the second phase (see Equation \ref{second-step}), showing the locations that are accessible within the threshold travel time from each demand site. Because of its properties, the 2SFCA approach has not only been widely used in spatial accessibility research, but it has also developed in multiple follow-ups (the so-called 2SFCA family), which complement each other to increase accuracy. The descendants' methodological breakthroughs are mainly differentiated according to three aspects (\cite{park_2021_a}), namely: (i) the different distance decay functions choosen \citep{luo2009, dai2010, mcgrail2014}, (ii) the different sizes of the catchment regions \citep{luo2012, tang2017}, and (iii) the customer preferences \citep{wan2012, delamater2013}.

In particular, the inclusion of customer's choices in the model led to a refinement of the 2SFCA, leading to the the Rational Agent Access Model (RAAM) (see \cite{saxon2020, saxon2021}), which incorporates patient feedback, it endogenizes the trade-off between travel time and congestion at the point of care, considers patients to seek care from any place and can accommodate also multiple transportation modes. In this model, we designate the fixed supply of ED at locations $l$ by $s_l$ and travel cost by $t_{kl}$. Patient demand, on the other hand, has an exact destination as well as an explicit origin (i.e. OD matrix \citep{lesage2008}), $p_{kl}$ indicates demand measured by the number of residents at location $k$. The cost of ED is calculated by adding the expenses of congestion and transport. The congestion cost is the observed inverse of the patients per provider ratio (PPR) at each location $l$ , adjusted to account for demand from all residential locations, and normalized by a factor $\rho$. The trip cost is just represented by the time $t_{kl}$ elapsed between an agent's residence $k$ and $l$ normalized by a parameter $\delta$. This parameter determines the cost or disutility of travel related to congestion. The entire cost for a resident of $k$ to obtain treatment at $l$ is therefore in equation \ref{raam}:

\begin{equation}\label{raam}
    RAAM(k,l) = \frac{ \sum \frac{p_{kl}}{s_l}}{\rho} + \frac{t_{k,l}}{\delta}
\end{equation}

This theory differs from 2SFCA (and its variants) in that it treats travel time as an explicit expense component rather than a weight on the availability of care from remote areas. RAAM's underlying idea is that information concerning $p_{kl}$ is already encoded in the cost function, the trip time matrix and the geographic distribution of supply and demand. Assuming that patients seek care in the most cost-effective location, the decision rule may be used to explain the choice of an agent at location $k$, this is obtained by minimizing RAAM for each location $l$. 
In other terms, patients select the place with the lowest cost constrained to the type of care they require. RAAM, like other basic economic models, assumes that patients have complete awareness of market costs and the absence of informative asymmetries.


\newpage

\section{Introducing new Accessibility-Reachability measures} \label{armat}

\subsection{The \textbf{AR} matrix}

Most of the methods discussed in Section 2 are focused on measuring the accessibility of population in different areas paying compartively less attention to the network of the supply locations. 

What we present in the present section is a new framework based on the gravity model and inspired by the spatial interaction modelling literature, but applied to stock data. In particular, the gravity model specification will be enriched with the addition of the catchment area as suggested by the 2SFCA method. Another innovation comes from the RAAM method, in that we consider only relevant agents costs constrained by the type of care needed (our analysis measure the accessibility only for white, yellow and green code i.e. only non critical conditions patients). Finally, our formalization will take in account simultaneously the interconnection between demand and supply networks.

In fact, reachability (defined as the ability to provide a certain service from the perspective of a health institution located at a given place) is equally significant for health planning purposes.

Based on the previous considerations, we present an alternative specification that allows for a combined examination of the accessibility of the various places from which the population originates and of the reachability of the supply sites. The two concepts are condensed into an object which we will refer to as the "accessibility - reachability matrix" which we designate with the symbol $\textbf{AR}$ matrix). 

In general, the  $\textbf{AR}$ matrix can be computed as follows. Let's consider a study area dividend into $l=1,\dots,L$ sub-areas (e. g. counties, neighbourhoods or simply squares in a lattice grid) where we observe $k=1,\dots,K$ supply infrastructures (e. g. hospitals), and let us further define $p_l \in \textbf{p}$ (\textbf{p} being a L-by-1 vector) as the population for each area $l$. Similarly, le $s_k \in \textbf{s}$ as a supply measure in each point $k$ (e. g. number of available beds in each hospital) (\textbf{s} being a K-by-1 vector).

With these definitions in mind $\textbf{C}$ can be defined as the (L-by-K) catchment matrix  characterized by the generic entry:
\begin{equation}
  c_{l, k} =
    \begin{cases}
      1 & \text{if $dist_1(l, k) \leq \tau$}\\
      0 & \text{otherwise}
    \end{cases}       
\end{equation}

$c_{l, k}\in \textbf {C}$, with $d(l, k) $ the distance between the centroid of the $l-th$ sub-area and the coordinate of the $k-th$ health structure, and with $\tau$ denoting the radius of the catchment area,
Furthermore, a second (L-by-K) distance matrix can also be computed characterized by the generic element

\begin{equation}
  d_{l, k} =
    \begin{cases}
      1 & \text{if $l=k$}\\
      dist_2(l,k)^{\gamma} & \text{otherwise}
    \end{cases}       
\end{equation}

$d_{l, k}\in \textbf{D}$ which considers an additional parameter $\gamma$ embodying a distance decay effect. The two distances $dist_1$ and $dist_2$ may differ in the unit of measurement (for example distance in km or travel time in minutes) or one may be variable over time (for example due to traffic or weather conditions).

We can then introduce the concept of the "extended population", incorporated in the p-by-1 vector $\textbf{p*}$, which relates to the population in each location augmented with the inclusion of shares of the population located in the neighboring areas which instist on the same infrastructure. This new element can be defined as follows:

\begin{equation}
 \textbf{p*} = \textbf{p}\textbf{C}\otimes\textbf{D}^T
\end{equation}

Each element of the $\textbf{AR}$ matrix, say $ar_{l,k}$, can then be computed as the ratio between the supply in location $k$ and the extended population in location $l$, multiplied by the corresponding elements of the two distance matrices previously introduced, that is: 

\begin{equation}
    ar_{l, k} = \frac{s_k}{p_l^*} c(l, k) d(l,k)^{\gamma}
\end{equation}

or, in matrix notation, 

\begin{equation}
    \textbf{AR} = \textbf{C} \otimes \textbf{D}^{T} \textbf{s}^{T} \overbar{p^{*}} 
\end{equation}

with $\overbar{p^{*}}$ defined as the vector with element $\overbar{p_l^{*}}$ = ($\overbar{p_l})^{-1}$.

Finally, the $\textbf{AR}$ matrix can be min-max standardized so as to remove the effects due to the different scales of measurement considered. 

Starting from the $\textbf{AR}$ matrix thus defined, we can introduce two useful indicators.

The first indicator refers to  the accessibility characterizing each area $l$, and it is defined as the row summation of the $\textbf{AR}$ matrix

\begin{equation}
   \textbf{A} = \textbf{AR} \iota_K
\end{equation}

with $\iota_K$ a K-by-1 unitary column vector and $A_l \in \textbf{A}$ its generic entry.

The second can be considered a "reachability" measure, say $R_k$, in each supplier location $k$ which is, conversely, calculated as the columns summation of the $\textbf{AR}$

\begin{equation}
   \textbf{R} = \iota_L^T\textbf{AR} 
\end{equation}

with $R_k \in \textbf{R}$ and $\iota_L^T$ a 1-by-L unitary row vector.

These two measures can be employed to the aim of defining impact measures as we will show in the next section.

\subsection{Impact analysis}

To this aim, let us first of all define a "supply-shock" as a sudden inaccessibility (e. g. due to saturation or unavailability) to a supply facility in one location occurring  \emph{coeteris paribus}. In this way we want to evaluate what happens to all infrastructures in the spatial system when one of them becames unavailable. Then the impact in each area caused by the shock is evaluated calculating the difference in the accessibility measure calculated before and after the shock. To formalize this calculation, let us define $ar_{l, k}^j \in \textbf{AR}^j$ as the $\textbf{AR}$ matrix resulting after having set to zero the supply at the generic location $j$ ($j=1,\dots,K$). Then the generic entry of the matrix can written as follows:

\begin{equation}
 ar_{l, k}^j = 
    \begin{cases}
      \frac{s_k}{p_l^*} c(l, k) d(l,k)^{\gamma} & \text{if $j \neq k$}\\
      0 & \text{if $j = k$}\\
    \end{cases} 
\end{equation}

We can then define the index  $\widehat{A_l}$  as:

\begin{equation}
\widehat{A_l} = min_{(j=1,K)}\left\{\sum_{k=1}^{K}ar_{lk}^j\right\}
\end{equation}

which corresponds to the accessibility measure in location $l$ when the most influential supply point $k$ is saturated.

Finally, since the impacts can be seen as the discrepancy between the observed accessibility and the accessibility after the removal of the supply, we can quantify them as the residuals of a linear  regression which explains the accessibility after the shock ($\widehat{A_l}$) as a function of the accessibility before the shock ($A_l$), that is: 

\begin{equation}
\textit{Impact} =  \widehat{A_l} - \alpha-\beta A_l 
\end{equation}

with $\alpha$ and $\beta$ the OLS estimates of the linear regression parameters.

\section{An empirical application: The case of emergency departments in Milan (Italy)}

\subsection{Data source and exploration}

The advent of big data is currently pushing the state-of-the-art statistical and econometric methodologies to the strain, not only due to the massive computing effort associated with the enormous volume and the speed with which data accumulate, but also because of the range of sources through which data is acquired \citep{arbia_2021_spatial}.
Most of these data are freely available, such as administrative documents or unstructured data derived from unexplored sources (such as crowdsourcing, web platforms, social media), and they are almost invariably geocoded. We will refer to all this variety of data as to the Spatial Web and Open Reliable Data (SWORD) which represent a potential massive source of knowledge that is of invaluable help in describing, monitoring, and forecasting spatial phenomena.
This is especially true for Big Data in healthcare, which includes massive collected data from electronic patient records (EPR) as well as streams of real-time geo-located health data acquired both with personal wearable devices and through public healthcare institutions, such as, for instance, Emergency Departments \citep{carr2009} or Pharmacies \citep{kostkova_2016_who} and many others.

Web scraping is the technique used for obtaining unstructured data from static or dynamic internet web pages and storing it in an organized manner. This procedure recently gained a lot of popularity because of the velocity and volumes through which data flows into the internet and the amount of valuable information that can enrich the existing data. 

In the present application, we used Web Scraping to download data from Emergency ward’s web pages using the L 15 web application provided by Regione Lombardia.
Although the collection of data has been carried out continuously every 5 minutes since November 2021 in the 115 ED of the region Lombardy (currently amounting to about 5 millions of data), for the case study reported in this section we have analyzed the data collected in the 18 ED of Milan observed on March 16, 2022 in three different moments of time at 3am, 9am and 3pm.
EDs accesses data are shown on a map along with their core specialities and general info. 

To quantify the supply side we considered the following variables:

(i) the number of patients waiting or in charge, split by triage code.

(ii) latitude and longitude of the ED and

(iii) data extraction time. 

On the other hand, from the demand side we considered the following variables:

(iv) population in each of the 1km-by-1km squares of a regular lattice grid overlapped on the Milan area.

(v) distance from ED coordinates to the centroids of the squares (determined under the assumption, that population in the grid is evenly spread as it is assumed by \citep{GEOSTATG56:online}).

The supply definition for the ED services is derived as the difference between the estimated ED maximum capacity (that is the maximum number of registered patients in charge for the whole time sequence per ED), and the current number of patients in charge.
Instead of considering the estimated maximum capacity one could argue that the exact number of beds might better express the notion of bed supply. However, unfortunately, these data are not available since the number of beds in italian EDs are not fixed (Decreto Ministeriale 2 aprile 2015 n. 70), and they can only be inferred based on the first decile of the population on the pertaining ED area. This led us to implicitly assume this number, relying on the assumption that the maximum capacity registered along the time sequence can well express an ED saturation scenario where all beds are occupied by patients in charge.

A further constraint to this new accessibility measure is represented by the target individuals upon which this measure is designed. Patients considered in this analysis are only those that are not in critical conditions (i.e. white code, green code, yellow code). As a matter of fact the  ED location to which the red coded patient travels to is usually determined by a range of factors which are either not known or are not rational ( i.e. hospital nearest location; ambulance owner, etc.) as discussed in \cite{saxon2020}. 


Moreover studies have been conducted on the correlation of max ED capacity (over the whole time sequence) and the estimated ED beds capacity inferred from neighbouring population areas. Results have shown a strong positive correlation (i.e. 0.69)  suggesting that our supply choice can be considered as an unbiased estimator of actual ED capacity.



\subsection{Data collection software}

\begin{figure}[t]
    \includegraphics[width=12cm]{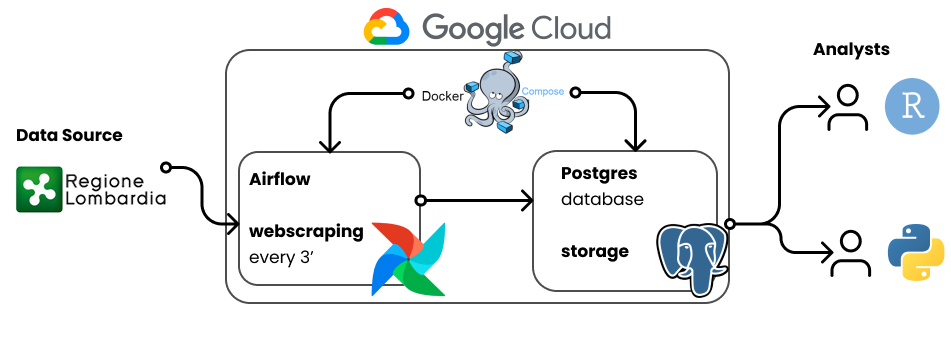}
    \caption{Data Collection Workflow for which Airflow schedules the extraction from the source i.e. L15 Lombardia web application. Data extracted as such is immediately stored into a relational PostgreSQL database. Docker Compose manages software services which are hosted on a Google Cloud virtual machine.}
    \label{fig:architecture}
    \centering
\end{figure}

The codes for data extraction are compiled in Python \citep{vanrossum_2009_python} and it is a combination of open source technologies and frameworks.
Apache Airflow is a data pipeline framework, which allows to schedule jobs (e.g. software executions) with a predefined chronological order.
Airflow carries out the data collection through scraping procedures and then stores the information into a SQL database i.e. PostgreSQL a sophisticated, open source object-relational database system that has been actively developed for over 30 years and it is recognized for dependability, feature robustness, and speed.
The data flows are then orchestrated through Docker \citep{merkel2014docker}, an open source technology that generates from scratch dedicated lightweight software environment (i. e. containers) that are portable and minimal in terms of dependencies. Moreover Docker-compose, which is an extension of Docker, is able to orchestrate services making it possible for diverse technologies to interoperate and to exchange data. Finally, a Virtual Machine (i.e. through the Google Cloud Engine paid service) hosts the aforementioned application and provides the scalable computational power needed to run the software. Figure \ref{fig:architecture} sketches the architecture that inspires the software presented.

\subsection{Calculation of the AR matrix}
\begin{figure}
    \includegraphics[width=\textwidth]{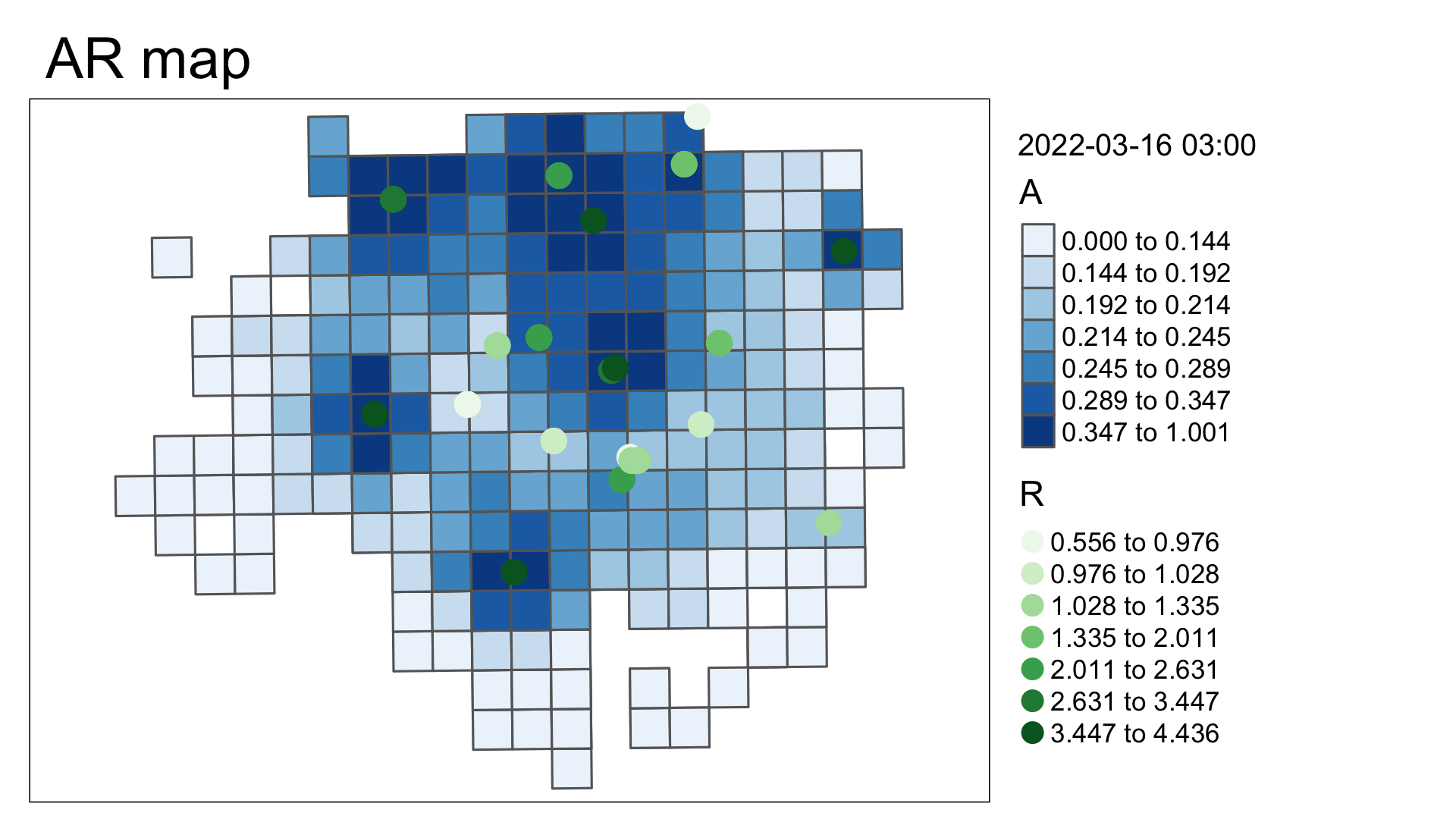}
    \label{fig:ar_3}
    \centering
    \caption{Map of Accessibility-reachability measures at 3am}
\end{figure}

\begin{figure}
    \includegraphics[width=\textwidth]{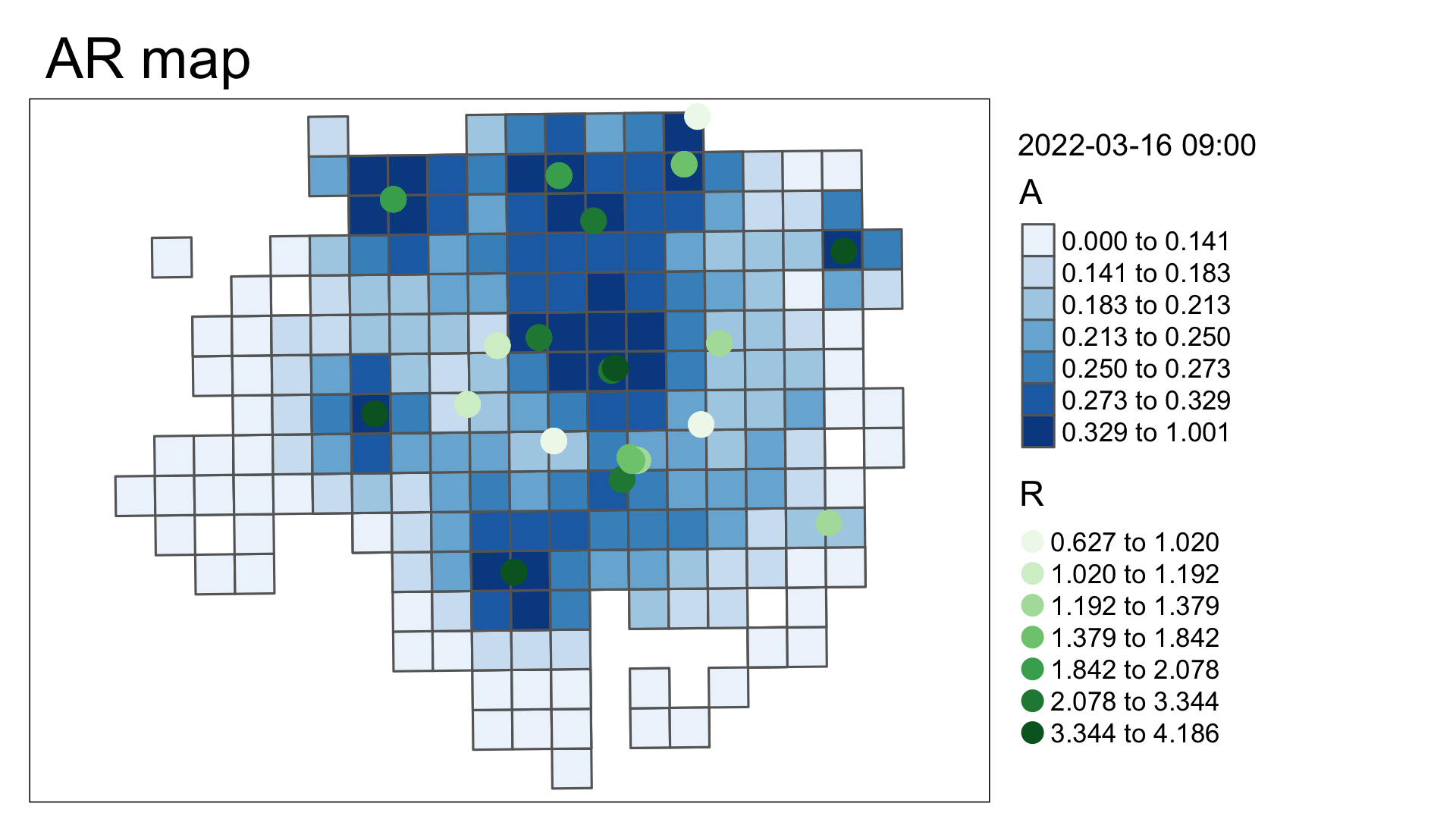}
    \label{fig:ar_4}
    \centering
    \caption{Map of Accessibility-reachability measures at 9am}
\end{figure}

In Section 3 we presented the $\textbf{AR}$  matrix, a measure that allows the combined examination of the accessibility of the various places from which the population originates and the reachability of the supply sites.
In contrast with the traditional methods presented in Section 2, our measure of accessibility-reachability concentrates on cooperative and non-cooperative behaviours (in the economics terminology as "economies and diseconomies of scales", see \cite{stigler58}) among EDs based on their capacity and their pairwise distances.
In our case study, we set the parameter $\gamma=-2$ in Equations 5 and 7 and we set the radious for the catchment area $\tau=5$ (projected euclidean distance in kilometers).
Figure 2, 3 and 4 show the maps of Accessibility-Reachability in the quadrat cells that constitute our study area observed in three different moments of time, namely 3am (Figure 2), 9am (Figure 3) and 3pm (Figure 4).
The intensity of the blue color of the cells indicates their level of accessibility to emergency departments, while the intensity of the green color associated to each hospital measures its reachability. The darker the color, the better it is in terms of health services accessibility  and reachability.
In the two maps observed in the morning (Figures \ref{fig:ar_3} and \ref{fig:ar_4}) the AR maps are very similar displaying a higher accessibility in  the central cells of Milan due to the greater concentration of EDs. Furthermore, in terms of reachability, EDs with low values are positioned in the center of the city while the opposite situation is observed for locations that are located at the border of the map.
In contrast with these empirical evidences, in the afternoon observations (Figure 4), accessibility drops significantly with the central cells experiencing the maximum accessibility at the values that were only the minimum in the morning (0.4).

\begin{figure}
    \includegraphics[width=\textwidth]{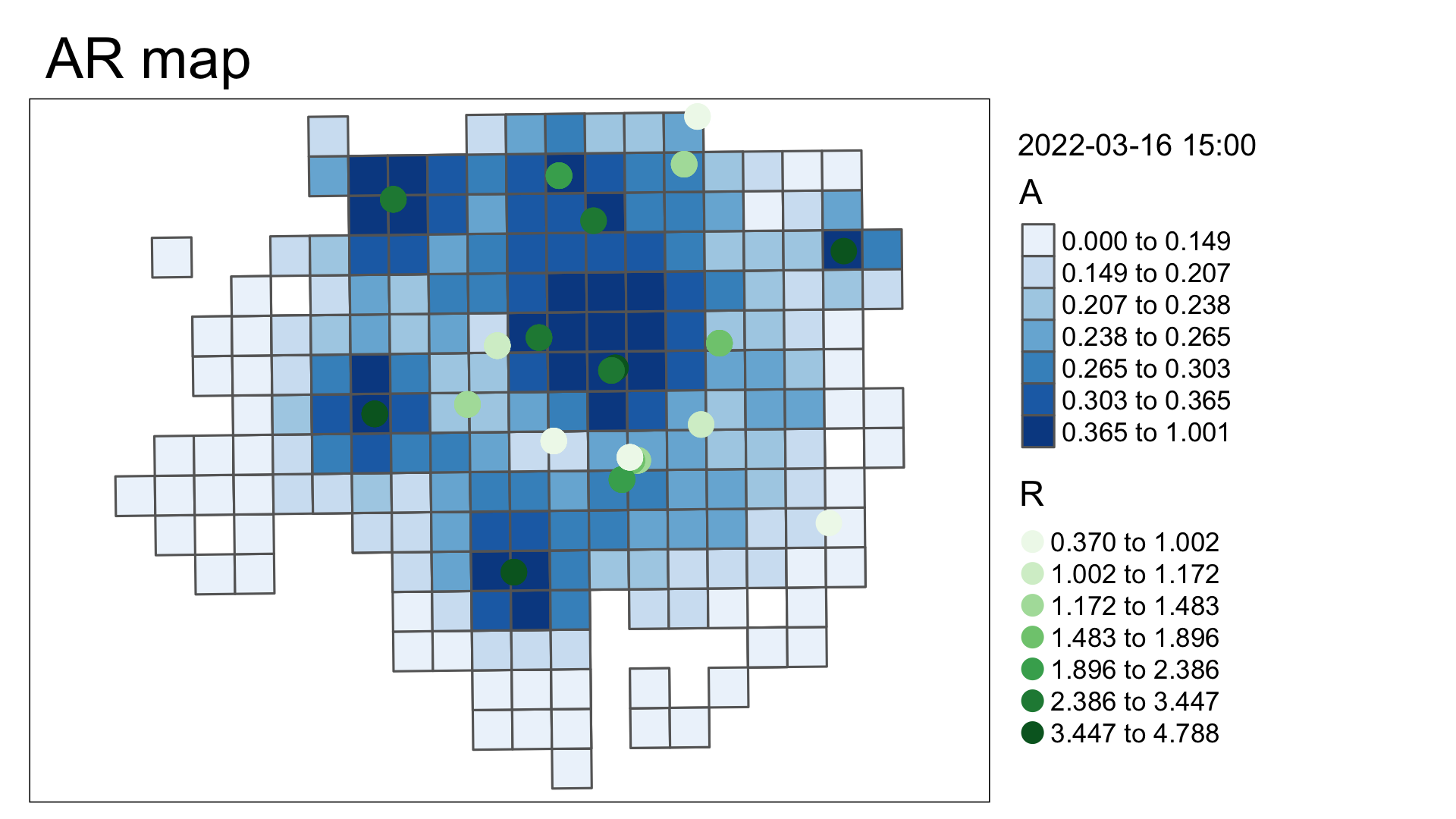}
    \label{fig:ar_5}
    \centering
    \caption{Map of Accessibility-reachability measures at 3pm}
\end{figure}

\subsection{Calculation of the impacts}

\begin{figure}
    \includegraphics[width=\textwidth]{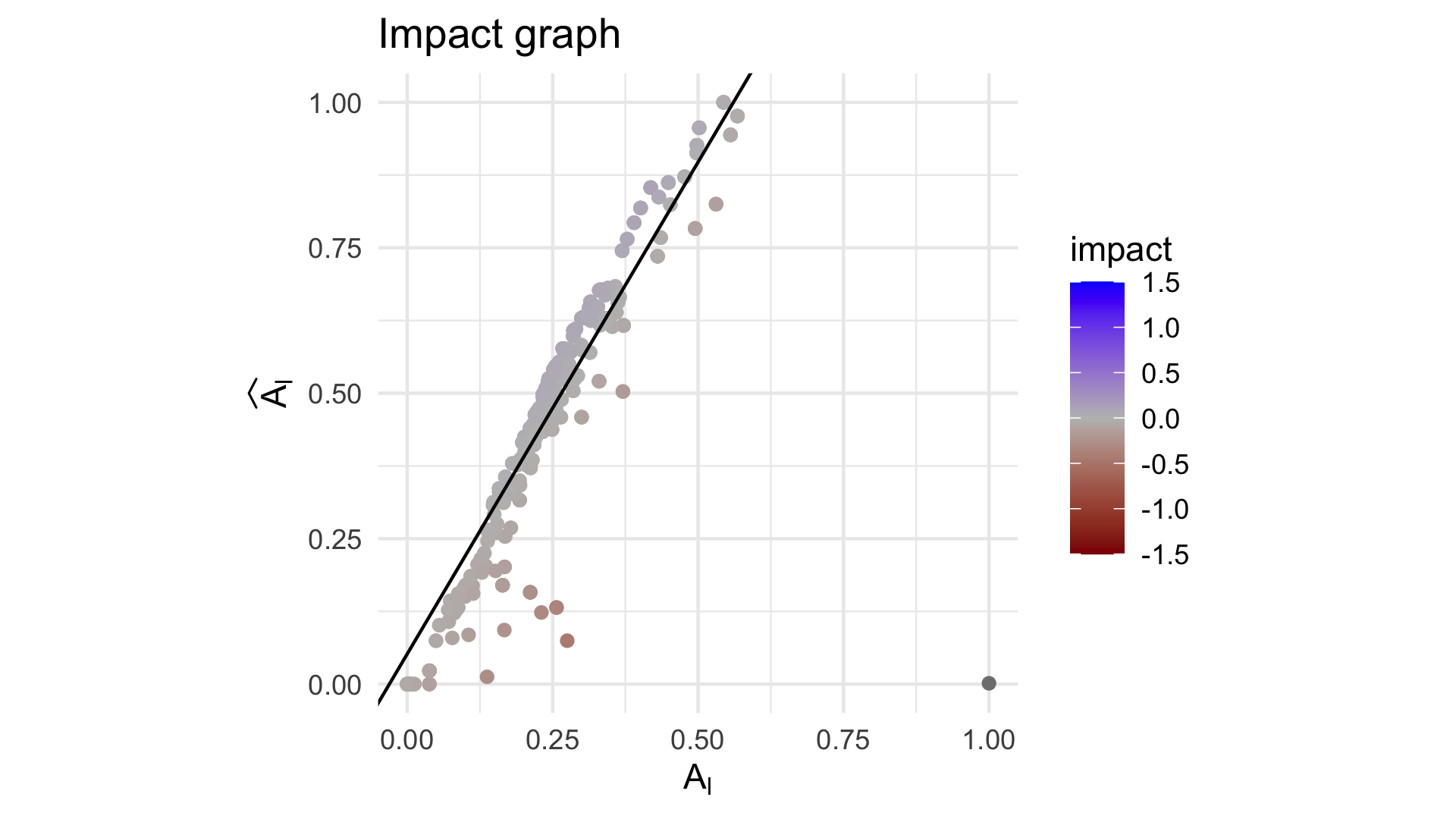}
    \caption{Impact graph}
    \label{fig:impact}
    \centering
\end{figure}

The impact analysis aims at measuring the effects of the saturation of a single ED on the supply network on the entire Emergency system in Milan. In the event of a shock, it is possible to define the inaccessibility (due to saturation of or unavailability) of a
single ED all other factors being constant.
Figure \ref{fig:impact} displays the effect of accessibility in case of overcrowding of one EDs.
As described in Section 3, in Figure \ref{fig:impact} we report on the vertical axis the variable $\widehat{A_l}$ (that represents the minimum accessibility observed when the hospital with the greatest influence has a supply equal to zero) and on the horizontal axis the $A_l$ variable, that is the (standardized) accessibility before the shocks.

By looking at Figure \ref{fig:impact}, when accessibility is high, but its minimum value is low (red dots), then all of the remaining accessibility depends on a single ED. If this emergency ward with greatest influence becomes saturated (or closed), the previously accessibility measured is lost. Because of this, the health services demand is poured out over alternative supplier.
The opposite situation occurs when both $A_l$ and $\widehat{A_l}$ are very high (blue dots).
The blue cells are positioned equidistant from EDs. In the event that one of the hospitals in the area is no longer available due to saturation, there would be no slow repercussions in accessibility.
The following maps show the impact that the closure of a hospital have in different moments of time (3am; 9am; 15pm).
The red cells refer to those that would suffer the most from ED saturation.
Indeed, for an ED it is not enough to be accessible, but it is also necessary to be reachable.
If the reachability of the closest ED goes to 0 and there are no alternatives then accessibility collapses to zero. In contrast, if there are other neighbors available it is likely that the accessibility remains constant.
Our results clearly display that more isolated EDs suffers more when the closest ED alternative is suddenly saturated in the newtwork in contrast with EDs located in middle and in N-E part of the surface. This is also true where population in grids is considerably high.
It is interesting to compare the impact during the three different time slots. At 3am and 9am the impacts are very similar: the reduction of accessibility is located in heterogeneity way. In the afternoon we observe the opposite whit the accessibility collapsing especially at the Milan's city limits.

\begin{figure}
    \includegraphics[width=\textwidth]{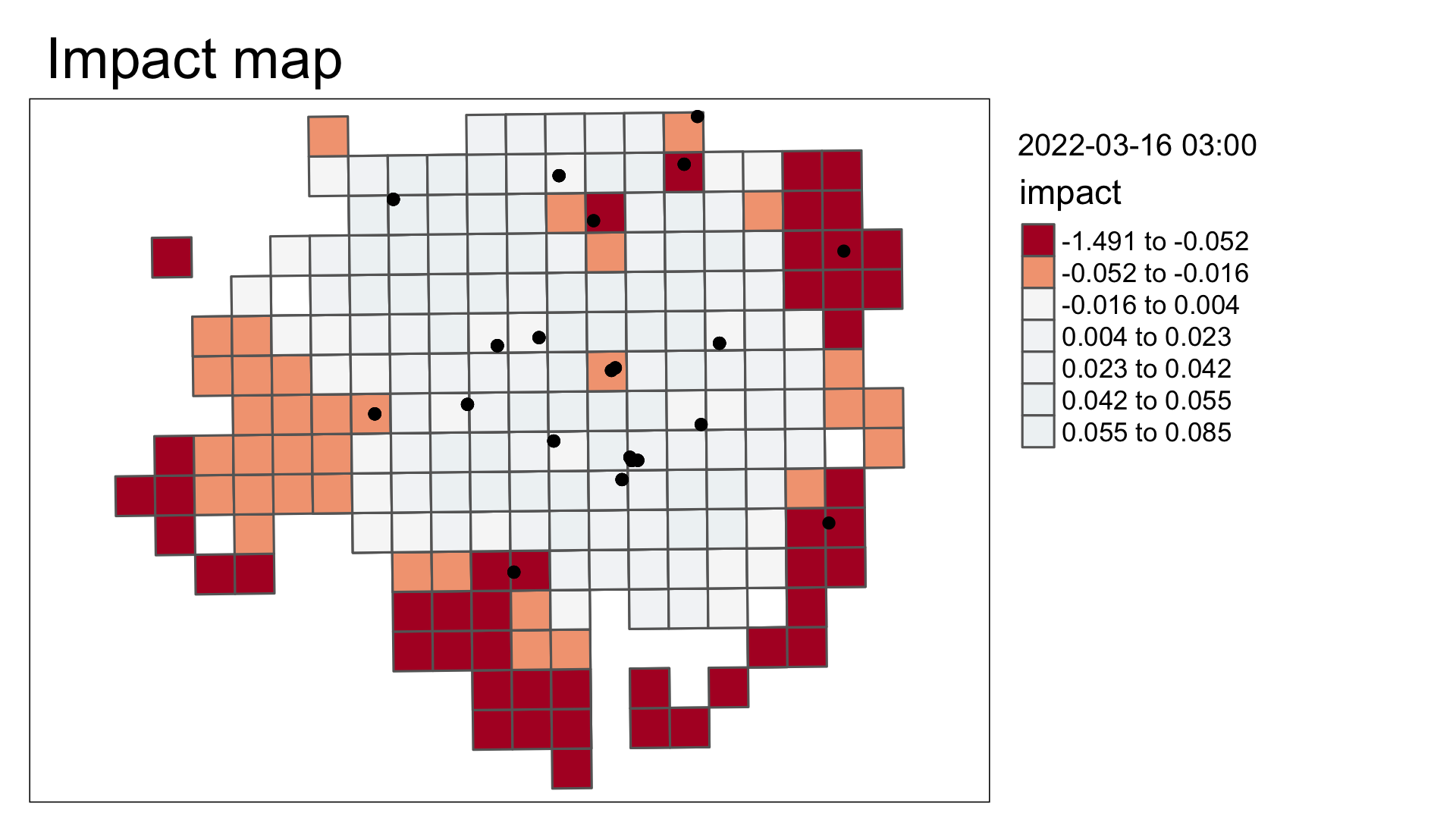}
    \caption{Impact map}
    \label{fig:impact_1}
    \centering
\end{figure}

\begin{figure}
    \includegraphics[width=\textwidth]{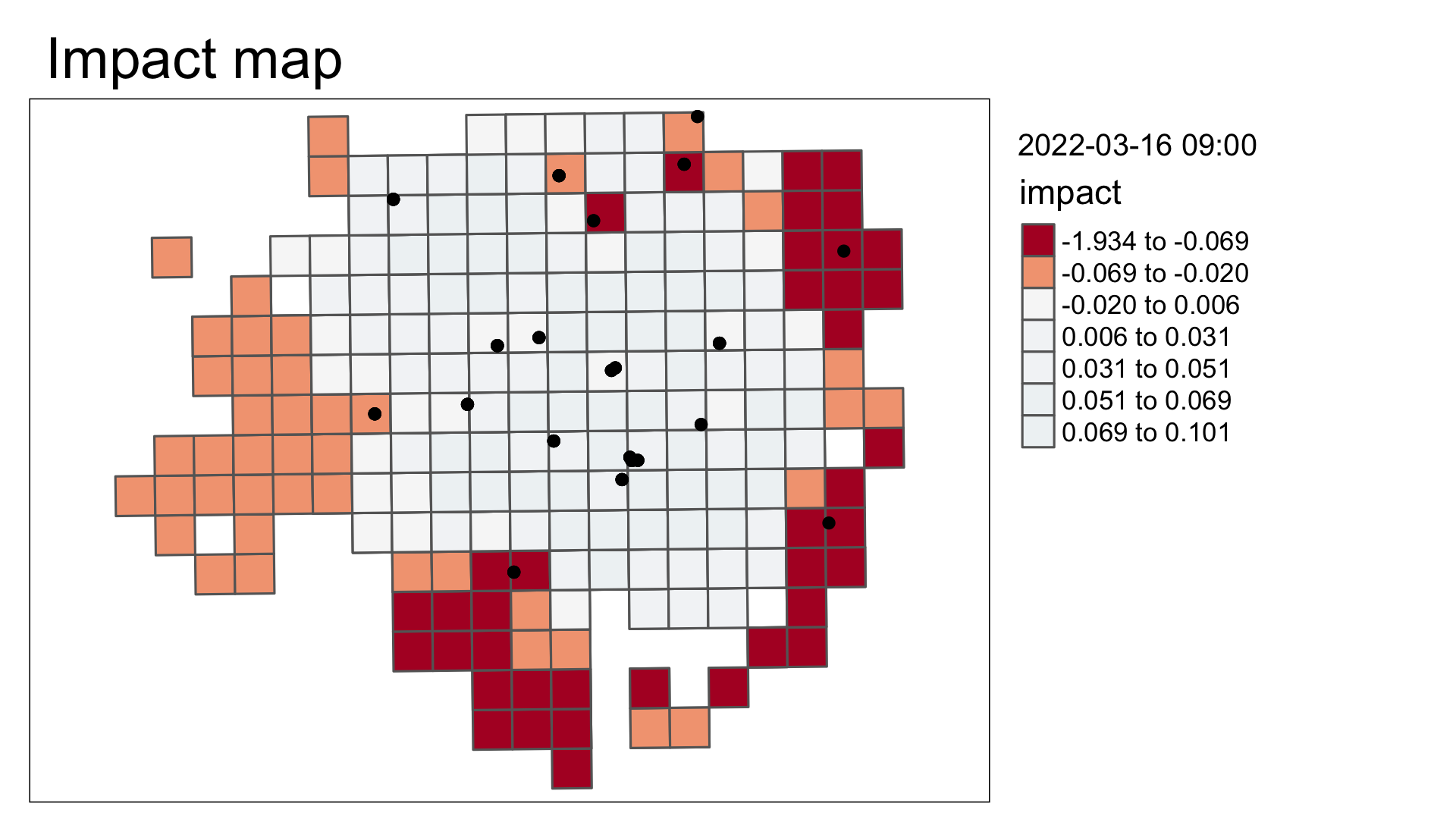}
    \caption{Impact map}
    \label{fig:impact_2}
    \centering
\end{figure}

\begin{figure}
    \includegraphics[width=\textwidth]{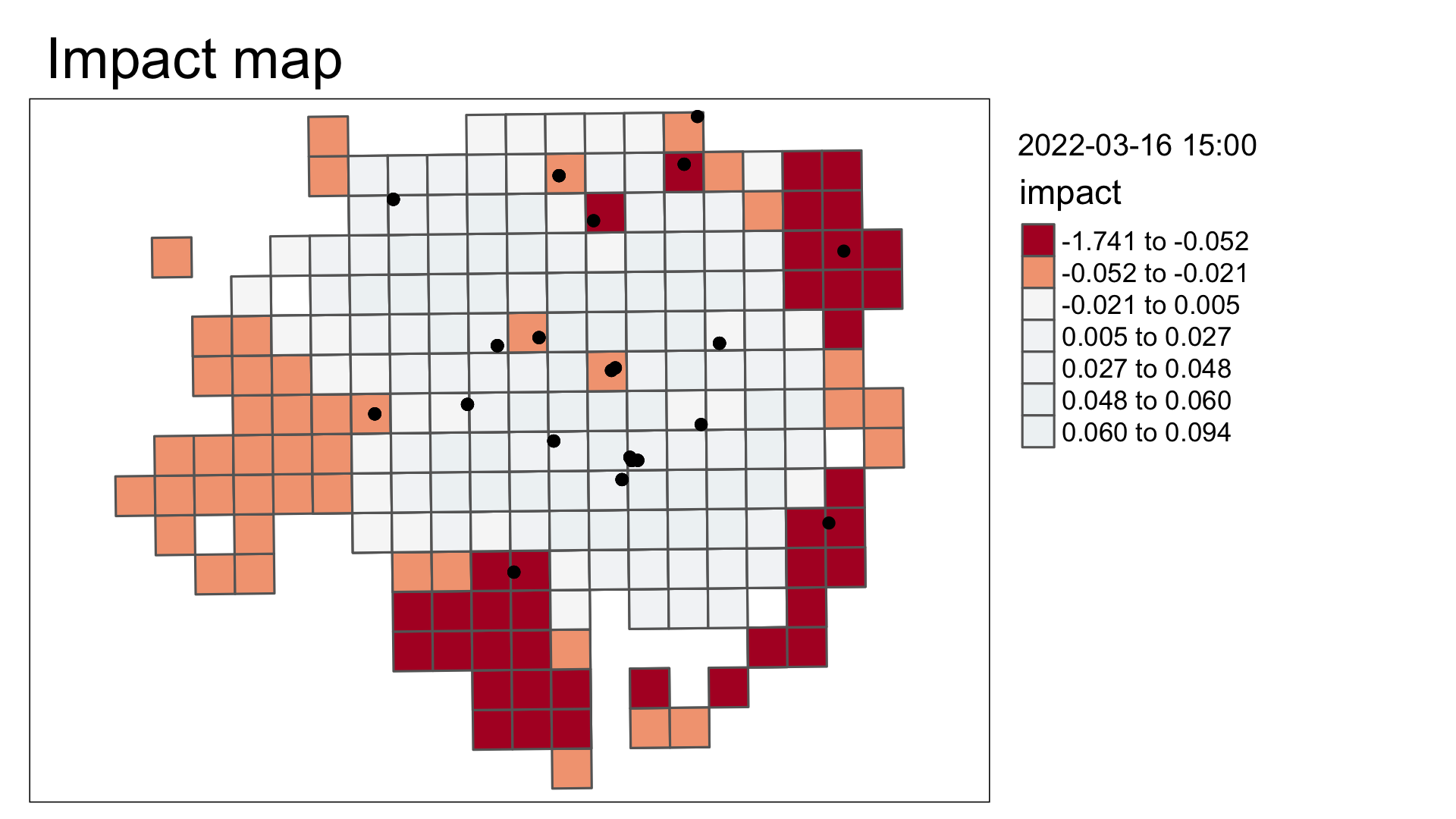}
    \label{fig:impact_3}
    \centering
    \caption{Impact map}
\end{figure}

\subsection{Testing intra-week accessibility differentials}

In the literature we find several studies that analyze 'weekend effect' on Emergency department. In this literature ED are compared in terms of appropriateness utilization, health outcomes and operational efficiency during weekends compared with what happens in the weekdays. \citep{schoenfeld2010weekend, duvald2018linking, wiedermann2018short}.
In the preset context we use the t-test to compare the accessibility in each area of the map between the weekend and the weekdays for two weeks ranging from March 7, 2022 to March 20, 2022. 
During the weekends General Practitioners (GPs) (those providing the first point of contact for patients) are not active and, as a consequence, in case of emergency, patients have to use EDs or Fist-Aid stations (called in Italy Guardia Medica).
For this reason, a mean hypothesis testing conducted through the Student's t-test, allows us to understand if there are significant differences between the number of people who refer to EDs during weekends with respect to weekdays.

The statistics of the twenty-first century is characterized by the presence of big data and thus by the need to carry out thousands of simultaneous hypothesis testing procedures \citep{Efron2016}. However, most of the classical hypothesis tests was built to be involved in just one experiment at time, so it is necessary to perform a procedure to correct the errors due to the multiple testing.

The simplest procedure is the so called Bonferroni test, categorized in the experiment-wise or family-wise error rate (FWER) approach.  Such test is the most conservative because it controls the probability that a true null hypothesis is incorrectly rejected among all possible hypotheses. Different methods were proposed to solve this issue with a stepwise procedure as suggested , e. g., in \cite{holm1979simple}, \cite{hommel1988stagewise} and \cite{hochberg1988sharper} among the many others.

\begin{figure}
    \includegraphics[width=\textwidth]{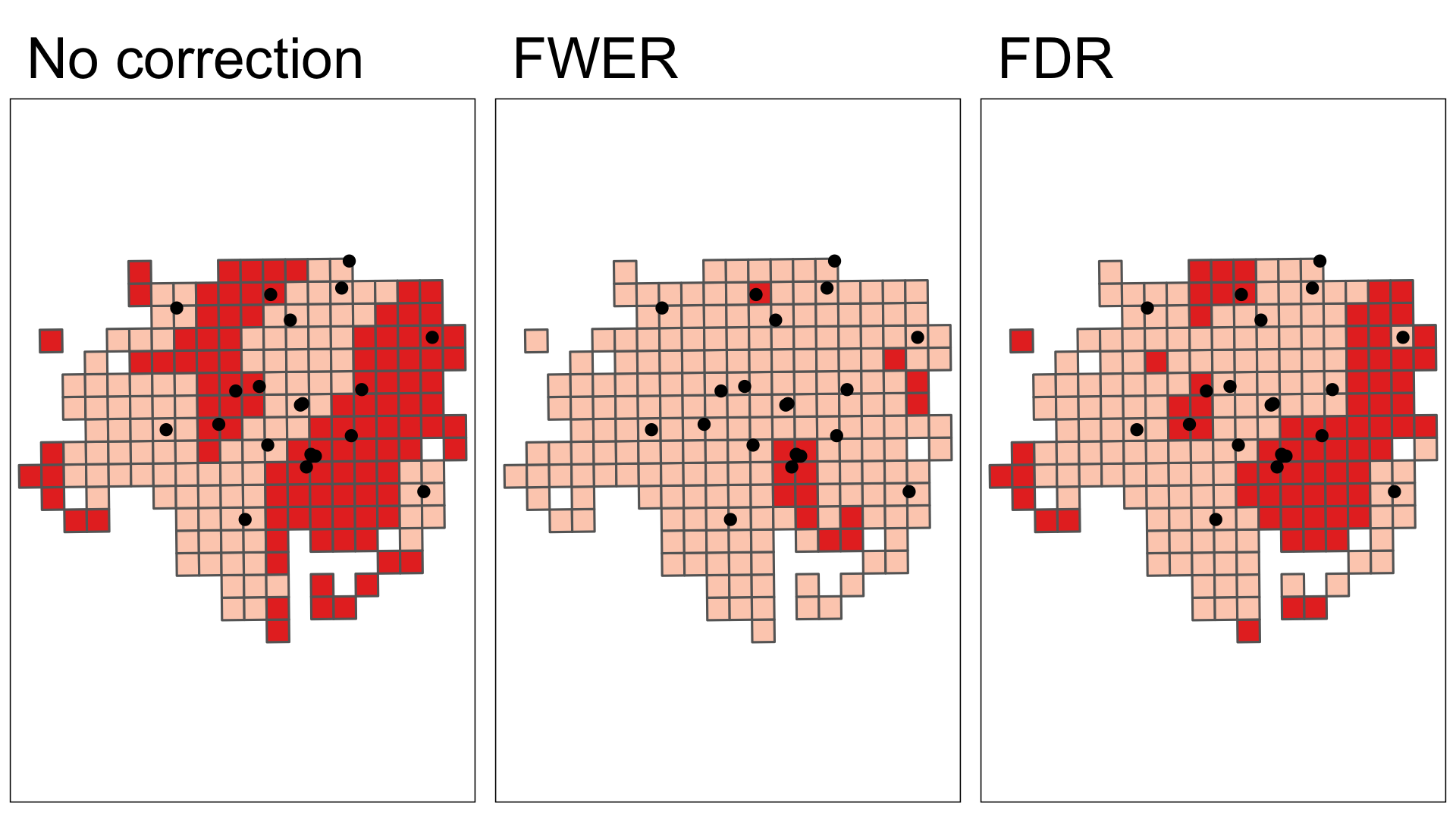}
    \caption{Areas (red) in which there has been a significant decrease in accessibility during the weekend considering the time slot between 7pm and 8pm with no correction, Bonferroni correction (FWER) and
    Benjamini-Hochberg (FDR).}
    \label{fig:ttest}
    \centering
\end{figure}

An alternative approach (initially proposed by \cite{Benjamini1995}) is to track the false discovery rate (FDR) defined as the expected proportion of type I errors among the rejections. This approach is particularly useful in exploratory data analysis in a big data context. In this application it is crucial to have the possibility to detect as many significant locations as possible in order to have the most accurate representation of the difference in accessibility observed between areas. The superiority of the FDR methods over FWER was proved using simulations \citep{williams1999controlling}. \cite{CaldasdeCastro2006} found that the original FDR specification, which assume independence between tests, provide better results even over the Bonferroni test with a spatial correction. For this reason, the following comparison will consider only the methods for independent testing, shifting the emphasis on spatial dependence to some future work.
These findings are confirmed also in our experiment: in Figure \ref{fig:ttest}, we highlight the areas where we observe significant decrease in accessibility during the weekend considering the time slot between 7pm and 8pm.

The test without any correction shows that out of 211 areas considered in the map 106 display a significant reduction in accessibility. This value contains false positives due to multiple testing. Indeed the more conservative Bonferroni correction (FWER) reduces these areas to only 14, while using the False Discovery Rate (Benjamini-Hochberg) we observe a significant decrease in 74 areas. Considering these results seems that the Benjamini-Hochberg test performs a correction that fits best for the application in hand. However, more investigations are needed in this area. 

From the results of the t-tests we can affirm that there is a significant difference of accessibility to EDs on weekdays and weekends. During the weekend, accessibility is most likely reduced due to the absence of GP's and to a wider of ED's.

\newpage

\section{Conclusions and research priorities}

Within the general literature of accessibility to health structures, the novelty of this paper is twofold.

First of all, we show how we can extract fresh open near-real-time geolocated data that can be sent to database where they are properly stored, and ultimately freely distributed and a new framework to web scraped data leveraging almost exclusively open source software is proposed.
In this respect, our experiment also shows how existing databases (either open or private) may well be enriched with unexploited open data acquired through non traditional sources.

Secondly, this paper introduces a new accessibility measure based not only demand, but also on overall supply network, involving Emergency Departments synergies and non-cooperative behaviours in EDs.

The approach presented here offers a number of practical possibilities to policy makers, including planning new ED, reinforcing or downgrading existing infrastructures and optimal resource allocations in emergency scenarios proving fast AR matrix computation which enables taking prompt decisions whenever new data become available.

After presenting the results related to the calculation of our proposed new measure of accessibility-reachability, in the empirical section of the paper, we studied how the saturation of a single ED on the network impacts the overall local ED system. Our results clearly show that the more isolated EDs suffer comparatively more than ED's clustered in center and in North-East part of Milan when the closest ED alternatives are suddenly saturated. This is also true when population density is higher. 

The empirical study then focused on testing if the presence of accessibility differences between weekdays and weekends. This analysis may help to verify the presence of significant differences in ED accessibility that can be traced back to activity of GP's.

The results obtained here could be extended in various directions.
First of all, the hypothesis testing approach presented here can be extended to various other situation, for instance to test the different accessibility between pre- and post-Covid 19 pandemics.
Secondly, future research could exploite space-time autoregressive models to forecast how accessibility evolves in time and space. Social and economical inequalities are also a broad field of investigation comparing richer and poorer accessibility areas.
Finally, a further interesting field of application could be the study of the impact of an unexpected event (i.e. earthquake, building collapse, fire, sport events etc.) located in certain areas, shifting from the supply shocks considered here to demand shocks.


\newpage

\clearpage
\bibliography{pronto-soccorso-methods}
\bibliographystyle{abbrvnat}
\end{document}